\documentclass[article,twocolumn,floatfix]{revtex4}
\usepackage{bm}
\usepackage{graphicx}
\usepackage{color}
\usepackage{amsmath}
\usepackage{amsfonts}
\newcommand{\mean}[1]{\ensuremath{\left \langle #1 \right \rangle}}
\newcommand{\be}{\begin{equation}}
\newcommand{\ee}{\end{equation}}
\newcommand{\bea}{\begin{eqnarray}}
\newcommand{\eea}{\end{eqnarray}}
\newcommand{\mvec}[1]{{\sf{#1}}}
\newcommand{\grad}{\vec{\nabla}}

\newcommand{\ddt}{\partial_t}
\newcommand{\ddd}[2]{\partial #1/\partial#2}
\newcommand{\Fn}[1]{\mvec{F}_{#1}}
\newcommand{\Frs}{\Fn{1}}
\newcommand{\Fo}{\Fn{0}}
\renewcommand{\Re}{{\rm Re}}
\renewcommand{\Im}{{\rm Im}}
\newcommand{\HR}{{\rm Hopf-Ra\~nada}\,}

\newcommand{\dd}{{\rm d}}
\newcommand{\half}{\tfrac12}
\newcommand{\EE}{\mathbb{E}^3}
\newcommand{\Cl}{{\cal C}_3}
\newcommand{\Clc}[1]{\overline{\mvec{#1}}}
\begin{document}

\title{The covariant description of  electric and magnetic field lines of null fields: application to \HR solutions}
\author{S.J. van Enk$^{1,2}$}
\affiliation{$^1$Physics Department and Oregon Center for Optics, University of Oregon, Eugene, OR 97403, USA}
\affiliation{$^2$Huygens Laboratory of Physics, University of Leiden,  2333 CA Leiden, The Netherlands}

\begin{abstract}
The concept of electric and magnetic field lines is intrinsically non-relativistic. Nonetheless, for certain types of fields satisfying certain {\em geometric} properties, field lines can be defined covariantly. 
More precisely, two Lorentz-invariant 2D surfaces in spacetime can be defined such that magnetic and electric field lines are determined, for any observer, by the intersection of those surfaces with spacelike hyperplanes.
An instance of this type of field is constituted  by the so-called \HR solutions of the source-free Maxwell equations, which have been studied because of their interesting {\em topological} properties, namely, linkage of their field lines.  
In order to describe both geometric and topological properties in a succinct manner, we
employ the tools of Geometric Algebra (aka Clifford Algebra) and use the Clebsch representation for the vector potential as well as the Euler representation for both magnetic and electric fields.
This description is easily made covariant, thus allowing us to define  electric and magnetic field lines covariantly in a compact geometric language. The definitions of field lines can be phrased in terms of 2D surfaces in space. We display those surfaces in different reference frames, showing how those surfaces change under Lorentz transformations while keeping their topological properties.
As a byproduct we also obtain  relations between optical helicity, optical chirality  and generalizations thereof, and their conservation laws.
\end{abstract}
\maketitle
\section{Introduction}
The concept of field lines in electromagnetics (EM) is interesting for various reasons.
Historically, Faraday considered field lines as the fundamental entities of
EM (see, e.g., Chapter~3 of Ref.~\cite{Faraday}). Moreover,  everyone's first acquaintance with the concept of magnetic field is, probably, through the simple experiment with iron filings that makes magnetic field lines around a bar magnet visible.
On the other hand, by the time one is an advanced student learning about Special Relativity (SR) and the role Maxwell's equations played in its development,  one typically does not hear much anymore about field lines. And indeed, the concept of field lines is highly non-relativistic for two reasons. First, electric and magnetic fields transform into each other under Lorentz transformations, and second, the points on a field line all have the same time coordinate. 

As it turns out,  one can define electric and magnetic field lines in a relativistically covariant manner only for certain types of solutions of Maxwell's equations. This has been known for magnetic field lines since a seminal paper by Newcomb \cite{Newcomb} (and as a demonstration of current interest in the covariant description of magnetic field lines, see \cite{pego}).
There he also discusses two related concepts that are logically prior to covariance of field lines, namely, {\em identity} of field lines and {\em motion} of field lines. For, in order to discuss what a given field line for one observer looks like to another observer,
one needs to be able to identify the field line in the first place. Similarly, even for one fixed observer, in order to be able to talk about the motion of a given field line one needs to identify which field lines at two different times are deemed to be the same. As Newcomb derived in Ref.~\cite{Newcomb}, magnetic field lines can be said to move with a (position and time-dependent) velocity $\vec{v}$ if and only if (employing units where $c=1$)
\be\label{Newcombv}
\grad\times(\vec{E}+\vec{v}\times\vec{B})=0.
\ee
He further derived that a covariant description of magnetic field lines exists if and only if
\be\label{EB0}
\vec{E}\cdot\vec{B}=0.
\ee
We will indicate how these conditions arise in Sections \ref{EuCl} and \ref{GAc}, respectively.
In the special case that (\ref{EB0}) is satisfied,
we can set $\vec{v}=\vec{E}\times\vec{B}/\vec{B}^2$ to satisfy (\ref{Newcombv}) identically. In the even more special case where $\vec{E}\cdot\vec{B}=0$ and $\vec{E}^2=\vec{B}^2$,
we have $|\vec{v}|=1$, i.e., the magnetic field lines move with the speed of light. Such fields are called ``null fields' and they are the subject of study in this paper.
 
Not all representations of the EM field are equally well suited for discussing and making covariant the concept of field lines. For example, the standard representation of the magnetic field consists of specifying $\vec{B}(\vec{r},t)$ as a function of position $\vec{r}$ and time $t$. The magnetic field line at time $t_0$  passing through a given point $\vec{r}_0$ is then a parametrized curve $\vec{f}(\lambda)$ found, in principle, by solving
the differential equation
\be
\frac{\dd \vec{f}(\lambda)}{\dd\lambda}=\frac{\vec{B}(\vec{f}(\lambda),t_0)}{|\vec{B}(\vec{f}(\lambda),t_0)|}
\ee
with ``initial'' condition $\vec{f}(0)=\vec{r}_0$.
Instead, we will make use of a more convenient representation, which does not require solving differential equations. When applied to the magnetic field, the representation goes under the name of Euler potentials \cite{Stern}. It leads naturally to a particular representation of the vector potential developed by Clebsch. This representation has several advantages in the present context: (i) it allows one to directly 
define magnetic field lines, (ii) it 
is easily made covariant for certain types of fields, (iii) it  shows that field lines can be linked only if the gradient of one of the potentials is singular, and (iv) in it, the magnetic helicity, a quantity known to contain topological information about the linkage of magnetic field lines \cite{Moffatt,Berger,Ranada1992b}, is compactly expressed.
These advantages will be exploited here as we are interested in displaying topological properties possessed by the (magnetic {\em and} electric) field lines of so-called \HR solutions 
\cite{Ranada1992, Ranada1997, Trueba1999, Irvine2008, Irvine2010}. These are solutions to the {\em free} Maxwell equations (without charges and currents).
In order to achieve the same 
compactness and simplicity  in describing their {\em electric} field lines, we make use of the dual  symmetry of the free Maxwell equations. In standard notation this symmetry transformation is  
\bea\label{dual}
\vec{E}&\mapsto&\cos\theta\vec{E}+\sin\theta\vec{B},\nonumber\\
\vec{B}&\mapsto&\cos\theta\vec{B}-\sin\theta\vec{E}.
\eea
This symmetry, and the geometric conditions a field has to satisfy in order to be able to define both electric and magnetic field lines, as well as covariance [not to mention Maxwell's equations], all turn out to be easily and much more compactly expressed within the formalism of Geometric Algebra (GA) \cite{Hestenesa,Hestenesb,Baylis,Doran,Lectures}. 
Because this formalism is not widely known (yet), we will start with 
standard non-covariant vector calculus and use it to define Euler potentials and field lines in Section \ref{EuCl}, and only then will we introduce and summarize the most relevant aspects of GA in Section \ref{GA}.
That formalism is used to describe EM noncovariantly in Section \ref{GAn} (with conservation laws of helicity and like quantities easily obtained), followed by the covariant description in Section \ref{GAc}. All that theory is applied to \HR solutions and a description of their field lines in Section \ref{Num}. In general, it is not easy to obtain Euler potentials for a given solution, but Ra\~nada
has provided explicit expressions for Euler potentials in his articles \cite{Ranada1992,Ranada1997} about the \HR solutions. In addition, those articles contain many deep insights, only some of which are exploited here.
In Section \ref{Bateman} we consider 
a much broader class of solutions to the free Maxwell equations, obtained long ago by Bateman \cite{Bateman}, for which a covariant description of field lines is in principle possible. We give these solutions in an elegant covariant form, without, however, being able to find the covariant Euler potentials.

\section{Euler potentials and the Clebsch representation}\label{EuCl}
\subsection{Euler potentials}
The Euler representation of the magnetic field amounts to writing (for historical and mathematical background, see \cite{Stern})
\be\label{Euler}
\vec{B}=\grad\alpha\times\grad \beta,
\ee
where $\alpha$ and $\beta$ are two scalar functions of position $\vec{r}$ and time $t$.
(Euler used this representation not for magnetic fields, of course, but for incompressible fluid flows, described by a flow velocity $\vec{v}_f$ with $\grad\cdot\vec{v}_f=0$.)
The advantages of representing $\vec{B}$ by Eq.~(\ref{Euler}) are that (i) the constraint 
\be\label{divB}
\grad\cdot\vec{B}=0
\ee 
holds automatically
if $\alpha$ and $\beta$ are sufficiently smooth functions of position, (ii) it conforms to the fact that in principle two scalar functions should suffice to express the three components of $\vec{B}$ subject to the single scalar constraint (\ref{divB}), and (iii) magnetic field lines at a given time $t$ are tangential to surfaces of constant $\alpha$ and to surfaces of constant $\beta$, and are thus determined by setting $\alpha$ and $\beta$ at time $t$ equal to constants, say, 
\bea\label{ab}
\alpha(\vec{r},t)&=&\alpha_0,\nonumber\\
\beta(\vec{r},t)&=&\beta_0.
\eea
Generically, the locus of points $\vec{r}$ satisfying these two constraints defines a 1D curve in 3D space, and that curve is a field line of $\vec{B}$. 

A disadvantage is that the Euler potentials are far from unique.
In fact, it is clear that any two functions $f(\alpha,\beta)$ and $g(\alpha,\beta$) determine the same magnetic field lines (and thereby the same $\vec{B}$ field) as long as the equations 
\bea
f(\alpha,\beta)&=&f(\alpha_0,\beta_0)\nonumber\\
g(\alpha,\beta)&=&g(\alpha_0,\beta_0),
\eea
have unique solutions for $\alpha$ and $\beta$. This will be the case whenever
the Jacobian for the transformation
$(\alpha,\beta)\mapsto(f,g)$
\be
J= \left| \begin{array}{cc}
    \ddd{f}{\alpha} & \ddd{g}{\alpha} \\ 
    \ddd{f}{\beta} & \ddd{g}{\beta} \\ 
  \end{array}\right|
\ee 
is never zero anywhere. More precisely, in that case we have
\be
\vec{B}=J^{-1}\grad f\times\grad g,
\ee
so that new Euler potentials $(f,g)$ are produced by any transformation $(\alpha,\beta)\mapsto(f,g)$ with $J\equiv 1$.
To demonstrate the disadvantage of the non-uniqueness, it suffices to note that, if only the Euler potentials were unique, then field lines could be given an identity straightforwardly through the use of (\ref{ab}), by identifying a field line at all times by the pair of constants $(\alpha_0,\beta_0)$. In turn this would immediately yield a (position- and time-dependent) velocity $\vec{v}$ of the magnetic field lines.
But as it is, one has to do a little bit more work to obtain a consistent definition of a field line velocity $\vec{v}$. Let us assume
that field lines can be assigned a field line velocity. In particular, assume there is a vector field $\vec{v}$ such that the time evolution of a pair of Euler potentials $(\alpha,\beta)$ satisfies
\bea
\ddt\alpha+\vec{v}\cdot\grad\alpha&=&0,\nonumber\\
\ddt\beta+\vec{v}\cdot\grad\beta&=&0.
\eea
Then, in order to obtain equations for the physical field $\vec{B}$ we multiply these two equations by the gradients of $\beta$ and $\alpha$, respectively, and then subtract the two. This yields \cite{Stern}
\be\label{Stern9}
\vec{v}\times\vec{B}=\ddt\alpha\grad\beta-\ddt\beta\grad\alpha.
\ee
We have not encountered the quantity appearing on the right-hand side yet, but the well-trained eye will recognize 
that it, together with $\grad\alpha\times\grad\beta$, will form a Lorentz covariant object (see Eq.~(\ref{FEuler}) below).
Taking the curl of (\ref{Stern9}) gives us an equation independent of the Euler representation:
\be
\grad\times(\vec{v}\times\vec{B})=\ddt\vec{B}.
\ee
Since we are interested in solutions to the source-free Maxwell equations, we can replace the term on the right-hand side
by $-\grad\times\vec{E}$, thus arriving at
\be
\grad\times(\vec{E}+\vec{v}\times\vec{B})=0,
\ee
which is the (necessary and sufficient) condition Newcomb \cite{Newcomb} found for a velocity $\vec{v}$ to be interpretable as a magnetic field line velocity.
\subsection{Clebsch representation and helicity}
Given Euler potentials $\alpha$ and $\beta$ it may seem straightforward now to define a vector potential $\vec{A}$  from which $\vec{B}$ can be derived as $\vec{B}=\grad\times\vec{A}$, namely by
\be\label{A}
\vec{A}=-\beta\grad\alpha.
\ee
There is, however, a subtlety associated with the magnetic helicity $H$ of the field, as has been discussed recently in, e.g., Ref.~\cite{Semenov}, as well as long ago in \cite{Dungey}, and by Ra\~nada in the context of the \HR solutions in \cite{Ranada1992b}. The magnetic helicity density  $h_m$ is defined as
\be\label{hm}
h_m=\vec{A}\cdot\vec{B},
\ee
in terms of which the magnetic helicity is given as an integral over all space
\be\label{Hm}
H_m=\iiint\!  h_m(\vec{r})\,\dd\vec{r}.
\ee
The form (\ref{A}) implies that $h_m=0$, and hence $H_m=0$. 
The next question is whether $H_m$ is gauge-invariant ($h_m$ is certainly not!). Under a  gauge transformation $\delta\vec{A}=\grad\psi$, we get, assuming we can apply Gauss's theorem,
\be
\delta H_m=\iiint(\grad\psi\cdot\vec{B})\,\dd \vec{r}=
\iint (\psi\vec{B}\cdot\vec{e}_n)\,\dd S
\ee
where the latter integral is a surface integral, with $\vec{e}_n$ denoting the unit vector normal to the surface.
For fields such that $|\vec{B}|\rightarrow 0$ sufficiently fast
for $r\rightarrow\infty$, with $r=|\vec{r}|$, we have that $\delta H_m=0$.
For the solutions we will consider the magnetic field does decay sufficiently fast, and so $H_m$ is gauge-invariant.
In particular, it may now seem that, in fact, $H_m=0$ for all such fields. One interesting aspect of the \HR solutions, however, is that we cannot so simply apply Gauss's theorem.
In particular, one of the Euler potentials---and we pick $\alpha$ here---must be multi-valued, and its gradient singular.
More precisely, $\alpha$ will be given in terms of
a particular complex function $\eta$ (to be defined below) as
\be
\alpha=\frac{1}{2\pi}\arctan\left(\frac{\Re( \eta)}{\Im( \eta)}\right).
\ee
This defines a multi-valued function, whose gradient is  singular in the locus of points $L$ where $\eta=0$, since
\be
\grad\alpha=\frac{\Im( \eta)\grad\Re( \eta)
-\Re( \eta)\grad\Im( \eta)}{2\pi|\eta|^2}.
\ee
Eq.~(\ref{A}) implies that $\vec{A}$ would be singular, too, at the locus $L$. However, we can choose a non-singular vector potential by adding a gauge-like term that cancels the singularity of the first term, viz.
\be
\vec{A}=-\beta\grad\alpha+\grad\Psi,
\ee
provided we pick $\Psi$ multivalued as well, according to
\be
\Psi=\beta(L)\alpha,
\ee
where $\beta(L)$ is the value (presumed unique)
attained by $\beta$ in the locus of points $L$.
If $\beta$ is determined by $|\eta|^2$ alone (which is true for the \HR solutions to be discussed below)
 $\beta$ takes on a unique value $\beta(L)$ in all points of the locus $L$. The helicity $H_m$ can now be nonzero
 by virtue of the multi-valued character of both $\Psi$ and $\alpha$  (and this singularity is, therefore, necessary in order to describe fields with linked field lines \cite{Moffatt,Berger}). In fact, we have
 \bea
  H_m&=&\iiint
 (\grad\alpha\times\grad\beta)\cdot\grad\Psi\,\dd\vec{r}\nonumber\\
&=&\beta(L) (\beta_{\max}-\beta_{\min})n^2(\alpha_{\max}-\alpha_{\min})^2,
 \label{Hmab}
 \eea
 where $n$ is an integer quantifying the type of multi-valuedness of $\alpha$ and $\Psi$: it counts the number of branches of the function $\alpha$. For the standard \HR solution we have $n=1$.
This expression makes manifest the precise relation between magnetic helicity and the Euler potentials.

\section{Geometric algebra}\label{GA}

Here we will briefly review the subject of Geometric Algebra (GA), also known as Clifford Algebra. Extensive introductions  can be found in articles by Hestenes
\cite{Hestenesa,Hestenesb}, who
has been advocating its use in physics over many years, as well as  in the two textbooks Refs.~\cite{Doran,Baylis}, and in a book of lecture notes \cite{Lectures}.

One could say that the idea of GA is to subsume both the dot product of (3D) vectors and their cross product under a single vector product that avoids certain 
shortcomings of the cross product, but it comes with a multitude of additional benefits, some of which will be made use of here.
The shortcomings of the cross product are that (i) it is not associative and (ii) the geometric notion that $\vec{a}\times\vec{b}$ is a vector pointing in the unique (up to a sign) direction perpendicular to the plane spanned by $\vec{a}$ and $\vec{b}$ does not generalize to higher dimensions. The vector product of GA {\em is} associative and its geometric meaning {\em does} generalize to any number of dimensions, for example, to the 4 dimensions of spacetime. Now it turns out that an elegant covariant description is possible using the GA of 3D Euclidean space $\EE$. That is, we do not need the GA of 4D Minkowski space, for reasons discussed in great detail in the book \cite{Baylis} and articles by Baylis \cite{Baylis1989,Baylis1996,Baylis2004} (see also below). We will, therefore, focus here on the GA of $\EE$. We denote that algebra by $\Cl$ (C for Clifford).

The vector product of two vectors
$\vec{a}$ and $\vec{b}$ 
is constructed to be associative and to obey distributive laws for addition and multiplication. An additional axiom is that the vector product of any vector with itself equal
 the length squared of the vector
 \be\label{axiom}
 \vec{a}\vec{a}=|\vec{a}|^2. 
 \ee 
The algebra is then built up from the set of all vectors by repeatedly taking sums and products of vectors.
We can in the end distinguish four (always one more than the dimension of the underlying vector space) different geometric types of basis elements. First, we have the standard vectors of $\EE$. Second, we encounter scalars, because of (\ref{axiom}). The third type arises when we consider the product of two linearly independent vectors, say, $\vec{a}$ and $\vec{b}$. Their product 
splits into a commuting symmetric scalar term $\vec{a}\cdot\vec{b}$ and an anti-commuting anti-symmetric ``bivector'' term (our third type of term) denoted by $\vec{a}\wedge\vec{b}$, the wedge product:
\be
\vec{a}\vec{b}=\half(\vec{a}\vec{b}+\vec{b}\vec{a})+\half(\vec{a}\vec{b}-\vec{b}\vec{a})=\vec{a}\cdot\vec{b}+\vec{a}\wedge\vec{b}.
\ee
The wedge product represents an oriented area for the plane spanned by the two vectors. Moreover, it generates rotations in that plane (both these meanings easily generalize to more dimensions).  
More precisely, if we use an orthonormal basis $\{\vec{e}_1,\vec{e}_2\}$
for a given plane, then any vector $\vec{c}\in\EE$ is rotated in the plane by an angle $\theta$ by the transformation
\be\label{rot}
\vec{c}\mapsto \vec{c}'=\mvec{R}\vec{c}\mvec{R}^{\dagger},
\ee
with
\be\label{R}
\mvec{R}=\exp(\half\theta\vec{e}_2\vec{e}_1)
\ee
and where the $\dagger$ operation reverses the order of vectors in any product.
One should note here  that $(\vec{e}_2\vec{e}_1)^2=-1$, so that we can also write
\be
\mvec{R}=\cos(\half\theta)
+\sin(\half\theta) \vec{e}_2\vec{e}_1.
\ee
This shows explicitly that $\mvec{R}$ is a sum of two different types of elements, a scalar and a bivector, and such elements of mixed type will always be denoted by sans serif symbols.

A product of three vectors can be expanded in vector terms and a ``trivector'' term, consisting of a product of three orthogonal vectors.
(This is the fourth and last type of term we encounter.) When we take those vectors to be unit vectors forming a right-handed frame, then we denote the resulting trivector by
\be
I=\vec{e}_1\vec{e}_2\vec{e}_3.
\ee
One can easily verify that this entity is the same for any right-handed set of orthogonal vectors. Moreover, it satisfies
\be
I^2=-1,
\ee
and it commutes with all bivectors and all vectors. It thus plays a very similar role as the standard imaginary unit $i$ (and it is denoted as such, by $i$, in many papers and books on GA; here we keep a different notation, to make sure we remember $I$ is, in fact, a trivector).
The geometric meaning of $I$ is that it represents an oriented volume, spanned by the three basis vectors (this geometric meaning makes it obvious there is only one such element in 3D, whereas in higher dimensions there exist multiple linearly independent trivectors). $I$ is a pseudoscalar as it changes sign under parity reversal.
Importantly, in 3D, the standard cross product is related to the wedge product by
\be\label{abI}
\vec{a}\wedge\vec{b}=I\vec{a}\times\vec{b}.
\ee
In words, the bivector $\vec{a}\wedge\vec{b}$ is dual to the cross product $\vec{a}\times\vec{b}$. More generally, multiplying a vector by $I$ yields a bivector; multiplying a bivector with $I$ yields a vector.

A general element $\mvec{M}$ of the algebra $\Cl$
is then a sum of the four types. Thanks to the general relation (\ref{abI}) we can write this sum as
\be\label{M}
\mvec{M}=a+b I +\vec{a}+I\vec{b},
\ee
where the terms represent the scalar, pseudoscalar, vector and bivector  parts, respectively, where $a,b$ are scalars, and $\vec{a},\vec{b}$ are ordinary vectors.

The sum of scalar and pseudoscalar parts is denoted by $\mean{\mvec{M}}_s$. Similarly, we denote the sum of vector and bivector parts by
$\mean{\mvec{M}}_v$.
A useful identity is
\be
\mean{\mvec{M}\mvec{N}}_s=\mean{\mvec{N}\mvec{M}}_s,
\ee
for any elements $\mvec{M}$ and $\mvec{N}$ in $\Cl$.
Note that the analogous relation for the vector part does not hold.
It is also useful to define ``real'' and ``imaginary'' parts of elements of $\Cl$ by
\be
\Re(\mvec{M})=(\mvec{M}+\mvec{M}^\dag)/2;
\,\,\,
\Im(\mvec{M})=I(\mvec{M}^\dag-\mvec{M})/2,
\ee
where the $\dagger$ operation, as mentioned above, reverses the order of vectors in a vector product; equivalently, it reverses the sign of $I$ in a decomposition like (\ref{M}).

Another involution, of great use in descriptions of relativity, is the Clifford conjugate, which reverses the directions of vectors and pseudovectors but leaves the scalar part the same. We denote the Clifford conjugate by a bar, like so
\be\label{Clc}
\Clc{M}=a+b I -\vec{a}-I\vec{b},
\ee
for $\mvec{M}$ given by (\ref{M}).

\section{Non-covariant description of electromagnetics within GA}\label{GAn}

\subsection{Preliminaries}
For the description of the EM field we will build on the ideas in the textbook \cite{Baylis}, going beyond its treatment in various aspects. This Section gives a non-covariant description of EM, the next presents the covariant version. The results in this Section pertain to general solutions of the free Maxwell equations, not just null fields, except in the very last subsection \ref{Hnull}.

It might be odd to begin this subsection by referring to the standard {\em covariant} description of EM, but this will explain why the non-covariant GA description is so elegant; moreover, it demonstrates that the description of EM in terms of the Riemann-Silberstein vector \cite{Birula}  (defined as $\vec{E}+i\vec{B}$) is  elegant for the simple reason that it is in fact the GA description with $I$ replaced by $i$ (see Eq.~(\ref{RS}) below). 

The standard covariant description of the EM field is in terms of the antisymmetric tensor $F_{\mu\nu}$ (with the indices running from 0 to 3). 
If we first define the following 4 elements (since they are not all of  the same type we use a sans serif symbol to denote them)
\be
\mvec{e}_0=1;\,\,\mvec{e}_{1,2,3}=\vec{e}_{1,2,3},
\ee
then
by defining
\be
\Frs=\half F^{\mu\nu}\mean{\mvec{e}_\mu\Clc{e}_\nu}_v
\ee
(using the usual Einstein convention of implied summation over repeated indices, and using the Clifford conjugate (\ref{Clc})) we get the central result \cite{Baylis}
\be\label{RS}
\Frs=\vec{E}+I\vec{B}.
\ee
Although the object on the left-hand side is covariant, the split into electric and magnetic fields is observer-dependent. In fact, $I$ is observer-dependent, because the three spatial unit vectors are.
\subsection{Source-free fields}
We now limit ourselves to a discussion of source-free fields.
For such fields we can introduce {\em transverse} (and thereby {\em gauge-invariant}) vector potentials by
\be\label{Fo}
\Fo=\vec{A}+I\vec{C},
\ee
such that
\be\label{FrsFo}
\Frs=\grad\Fo.
\ee
Indeed, when considering separately the four geometric types contained in this equation, we find that the scalar and pseudoscalar parts give $\grad\cdot\vec{A}=0$ and $\grad\cdot\vec{C}=0$, respectively.
The sign convention of (\ref{Fo}) is such that $\vec{E}=-\grad\times\vec{C}$. The reason for introducing $\vec{C}$ is so we can define an electric helicity. In analogy to
(\ref{hm}) we define the electric helicity density
\be
h_e=\vec{C}\cdot\vec{E},
\ee
and the electric helicity $H_e$ is then obtained by integrating this density over all space.

We can inductively define a whole hierarchy of fields and/or potentials by
\be
\Fn{n+1}=\grad\Fn{n},
\ee
for all integer $n$. For example,
\be
\Fn{2}=-\grad\times\vec{B}+I\grad\times\vec{E}.\ee
For free fields they all satisfy
\be
(\ddt+\grad)\Fn{n}=0.
\ee
All these equations are then invariant under the duality transformation (\ref{dual}), which is expressed simply as
\bea
\Fo\mapsto \exp(-I\theta)\Fo.
\eea
This transformation induces the transformation
\bea
\Fn{n}\mapsto \exp(-I\theta)\Fn{n}.
\eea
Having defined the dagger operation before as reversing the sign of $I$, we also have
\bea
\Fn{n}^\dagger\mapsto \exp(I\theta)\Fn{n}^\dagger.
\eea 
This makes it straightforward to construct quantities that are invariant under the duality transformation, namely as bilinear quantities containing one term $\Fn{n}$ and one term
$\Fn{m}^\dagger$ and possibly more factors that are field independent. Such quantities are discussed in the next subsection.
\subsection{Bilinear quantities}
 Here we give the relations between certain bilinear geometric algebraic quantities and
the more familiar bilinear vectorial quantities.
Since the elements of $\Cl$ do not commute in general, it is convenient to consider symmetric and anti-symmetric bilinear quantities.
For example, we have
\bea
\Frs^\dag\Frs+\Frs\Frs^\dag&=&2(\vec{E}^2+\vec{B}^2),\nonumber\\
\Frs^\dag\Frs-\Frs\Frs^\dag&=&-4\vec{E}\times\vec{B},\nonumber\\
\Re(\Frs^2)&=&\vec{E}^2-\vec{B}^2,\nonumber\\
\Im(\Frs^2)&=&
2I\vec{E}\cdot\vec{B}.
\eea
We can recognize the energy density of the field, the momentum density, and the Lagrangian density in the first three lines. The fourth line gives the quantity that should vanish if a covariant definition of electric and magnetic field lines is to be possible, according to \cite{Newcomb}.
Furthermore, we have
\bea
\label{helicity}
\Frs\Fo+\Fo\Frs&=&2(\vec{E}\cdot\vec{A}-\vec{B}\cdot\vec{C})
+\nonumber\\&&
2I(\vec{B}\cdot\vec{A}+
\vec{E}\cdot\vec{C})\\
\Frs\Fo-\Fo\Frs&=&2I(\vec{E}\times\vec{A}-\vec{B}\times\vec{C})
+\nonumber\\&&
-2(\vec{B}\times\vec{A}+
\vec{E}\times\vec{C})
\eea
These two definitions are not invariant under duality transformations.
Such non-invariant quantities are useful to distinguish quantities that are conserved because of some symmetry from those that are conserved only for particular solutions (see below). Moreover, since measurements on the free EM field may make use of electric charges, measurement can break the duality symmetry, and thus
it is not true that only duality-invariant quantities are physically relevant for free fields, as was argued in \cite{Barnett,Barnett2}.
A duality-invariant version of Eq. (\ref{helicity}) is
\bea\label{heln}
\Frs\Fo^\dag+\Fo^\dag\Frs&=&2(\vec{E}\cdot\vec{A}+\vec{B}\cdot\vec{C})
+\nonumber\\&&
2I(\vec{B}\cdot\vec{A}-
\vec{E}\cdot\vec{C})
\eea
In the second line the total (i.e., magnetic plus electric) helicity density appears as a pseudoscalar.
Combining similar relations yields the magnetic and electric  helicity densities separately:
\bea
\vec{B}\cdot\vec{A}&=&(\Im(\Frs)\Re(\Fo)+\Re(\Fo)\Im(\Frs))/2,\nonumber\\
\vec{E}\cdot\vec{C}&=&(\Re(\Frs)\Im(\Fo)+\Im(\Fo)\Re(\Frs))/2.\nonumber\\
\eea
We conclude this subsection by noting that duality invariance and its relation to helicity and the separation of the total angular momentum of light into spin and orbital parts has become the subject of very recent studies \cite{Cameron2012,Bliokh2012}.

\subsection{Conservation laws for (pseudo)scalar quantities}
We consider here in some detail how one may show within the GA formalism that a (pseudo)scalar quantity is conserved. Take helicity as an example. We use three 
handy maneuvers.
 The first is to explicitly take the scalar plus pseudoscalar part of 
\bea
\ddt\mean{\Frs\Fo^\dag}_s&=&\mean{\ddt(\Frs\Fo^\dag)}_s\nonumber\\
&=&-\mean{\grad\Frs \Fo^\dag}_s-\mean{\Frs(\grad \Fo)^\dag}_s\nonumber\\
&=&-\mean{\grad\Frs \Fo^\dag}_s-\mean{\Frs\Frs^\dag}_s.
\eea
The second maneuver is to note that 
\be
\mean{\grad(\Frs \Fo^\dag)}_s=\mean{\grad\mean{\Frs \Fo^\dagger}_v}_s,
\ee
because the gradient vector multiplying a scalar term can never give rise to a scalar term.
The gradient on the lhs acts on both $\Frs$ and $\Fo^\dag$. But because general elements of $\Cl$ do not commute, in order to calculate the result from the gradient acting on the second term, we first move it to the front before taking the derivative. That is,
the third maneuver is to write
\bea
\Frs \Fo^\dagger=-\Fo^\dagger\Frs+2\mean{\Frs \Fo^\dagger}_s,
\eea
and use this to rewrite
\bea
\mean{\grad(\Frs \Fo^\dagger)}_s&=&
\mean{\grad\Frs  \Fo^\dagger -\grad(\Fo^\dagger)\Frs}_s\nonumber\\
&=&\mean{\grad\Frs  \Fo^\dagger +\Frs^\dagger\Frs}_s,\nonumber\\
\eea
where in the last line we used
\be\label{grad}
\grad(\Fn{n}^\dag)=-(\grad\Fn{n})^\dag.
\ee
All this together yields
\bea
\ddt\mean{\Frs\Fo^\dag}_s
&=&-\grad\cdot \mean{\Frs \Fo^\dag}_v
+\mean{\Frs^\dag\Frs}_s\nonumber\\
&&-\mean{\Frs\Frs^\dag}_s\nonumber\\
&=&-\grad\cdot \mean{\Frs \Fo^\dag}_v.
\eea
This shows that the helicity is conserved for free fields, but so is
the quantity
\be
Q=\vec{E}\cdot\vec{A}+\vec{B}\cdot\vec{C}.
\ee
This quantity $Q$ is not that interesting, however, as it is just proportional to the time derivative of  $\vec{A}^2+\vec{C}^2$ (which we see when substituting the well known relation $\vec{E}=-\ddt\vec{A}$, as well as the less known but similar relation $\vec{B}=-\ddt\vec{C}$), which is itself a conserved quantity.
\subsubsection{Helicities and null conditions}\label{Hnull}
We now wish to consider the magnetic and electric helicities separately.
For this purpose it suffices now (given the above results) to consider
\bea\label{F1F0}
\ddt\mean{\Frs\Fo}_s&=&\frac12\mean{\ddt(\Frs\Fo+\Fo\Frs)}_s\nonumber\\
&=&-2\Frs^2-\grad\cdot\mean{\Frs\Fo}_v,
\eea
where above-mentioned maneuvers were used once again.
This shows that the magnetic and electric helicities are {\em not} individually conserved. But for solutions for which $\Frs^2=0$, they are individually conserved.
Solutions of Maxwell's equations with $\Frs^2=0$ are termed ``null solutions'' for obvious reasons, and they have been considered in various contexts both long ago and very recently \cite{Bateman,Besieris,Besieris2,Birula2} (see also Section \ref{Bateman}).
Now  taking the time derivative of $\Frs^2$ (which is identically zero for a null field, of course) we find
\be\label{F2}
\ddt\Frs^2=-2\mean{\Fn{2}\Frs}_s.
\ee
The quantity on the rhs is the {\em difference} between two higher-order helicities, not related to the field lines of $\vec{E}$ and $\vec{B}$, but to those of $\grad\times\vec{E}$ and
$\grad\times\vec{B}$.  Clearly, it vanishes for null fields. (The {\em sum} of those higher-order helicities  is, in fact, the optical chirality density, which has become of great interest recently \cite{Tang,Tang2,Hendry,Bliokh, Barnett,Barnett2,Coles}.)

We can now repeat the story: since the right-hand side must be zero, its time derivative vanishes as well. But that time derivative is now easily found by reference to (\ref{F1F0}), 
\bea
\ddt\mean{\Fn{2}\Frs}_s&=&\frac12\mean{\ddt(\Fn{2}\Frs+\Frs\Fn{2})}_s\nonumber\\
&=&-2\Fn{2}^2-\grad\cdot\mean{\Fn{2}\Frs}_v.
\eea
This way we encounter higher-order versions of the helicity differences, $\mean{\Fn{n+1}\Fn{n}}_s$, as well as higher-order versions of the null conditions, $\Fn{n}^2=0$, and they are related through conservation laws.
\section{Covariant description of electromagnetics within GA}\label{GAc}
\subsection{The Baylis method}
Special relativity can be compactly described in a covariant manner by using the spacetime algebra \cite{Hestenesb}, i.e., the GA constructed from spacetime vectors living in 4D Minkowski spacetime.
However, perhaps surprisingly, $\Cl$ can be used 
just as well for the same purpose. This has been demonstrated by Baylis and coauthors in several articles \cite{Baylis1989,Baylis1996,Baylis2004}. 
There is a subtle difference between the two descriptions, related to the fact that the spacetime algebra has twice as many basis elements as $\Cl$ (it is 16-dimensional whereas $\Cl$ is 8-dimensional). Physical quantities such as the (rest) mass of a particle or the proper time of a particle {\em in its own rest frame} play two roles: as the zeroth component of the energy-momentum four-vector or the spacetime position four-vector, respectively,  but also as the Lorentz-invariant length of the respective four-vectors (in any frame, of course). Whereas $\Cl$ uses a single representation for either of these two roles, spacetime algebra has two different representations, one for each role.
Here we follow Baylis' lead (see also his textbook \cite{Baylis}) and use the simpler algebra $\Cl$.

Any four-vector is represented by a ``paravector'', i.e., a real element of $\Cl$. That is, we can write a general
paravector as
\be
\mvec{p}=p_0+\vec{p},
\ee
with $p_0$ a scalar and $\vec{p}$ a 3D vector.
For example, the spacetime four-vector (as defined relative to some arbitrarily chosen fixed origin of Minkowski space) is represented by the paravector
\be
\mvec{r}=t+\vec{r}.
\ee
The Minkowski metric is obtained in  natural way
by defining the ``square length'' of a paravector to be
\be
\mvec{p}\Clc{p}=p_0^2-\vec{p}^2,
\ee
in terms of the Clifford conjugate (\ref{Clc}).
The minus sign here implies that another important example of a paravector, the gradient four-vector,  takes the form
\be
\partial=\ddt-\grad.
\ee
The scalar product between different paravectors $\mvec{p}$ and $\mvec{q}$ is defined as the scalar part
of the product $\mvec{p}\Clc{q}$, i.e.
\be
\mean{\mvec{p},\mvec{q}}=
\mean{\mvec{p}\Clc{q}}_s=\half(\mvec{p}\Clc{q}
+\mvec{q}\Clc{p}).
\ee 
A Lorentz transformation preserves, by definition, this scalar product of paravectors. It acts on paravectors as
\be
\mvec{p}\mapsto \mvec{p}'=\mvec{L}\mvec{p}\mvec{L}^\dagger,
\ee
where $\mvec{L}$ can be chosen to satisfy
\be\label{L}
\mvec{L}\Clc{L}=1.
\ee
For example, a pure boost is represented as
\be
\mvec{L}=\exp(\vec{w}/2),
\ee
where $\vec{w}$ is the so-called ``rapidity.''
Compare this to the description of rotations in space, (\ref{rot}) and (\ref{R}), which are, of course, also elements in the group of Lorentz transformations.
Using (\ref{L}) we also obtain the combined action of Lorentz transformations and involutions:
\bea
\Clc{p}&\mapsto& \Clc{L}^\dagger\Clc{p}\Clc{L},\nonumber\\
\mvec{p}^\dagger&\mapsto& \mvec{L}\mvec{p}^\dagger \mvec{L}^\dagger.
\eea 
Given the condition (\ref{L}) on Lorentz transformations,
one easily sees that alternating products of paravectors and Clifford conjugates of paravectors transform in a simple way. When there is an odd number of paravectors and conjugates, then the object transforms just as a paravector or as the Clifford conjugate of a paravector. When there is an even number of paravectors and conjugate terms in the alternating product, then it transforms slightly differently. For example, an object
$\mvec{O}=\mvec{p}\Clc{q}$ transforms as
\be
\mvec{O}\mapsto \mvec{O}'=\mvec{L}\mvec{O}\Clc{L}.
\ee
We can distinguish two contributions to $\mvec{O}=\mvec{p}\Clc{q}$:
\be
\mvec{O}=\mean{\mvec{p}\Clc{q}}_s+
\mean{\mvec{p}\Clc{q}}_v,
\ee
where the first term is just the Lorentz-invariant scalar product of $\mvec{p}$ and $\mvec{q}$, and for the second type of term (i.e., a sum of vector and pseudovector parts)
Baylis uses the name ``biparavector.''
\subsection{The EM field}
The EM field is described by a biparavector, namely
$\Frs$. The covariant form of the free Maxwell equations is really the same as before:
\be
\overline{\partial}\Frs=0.
\ee
We can introduce a gauge potential by
\be
\Frs=\mean{\partial \Clc{A}}_v.
\ee
Now here is an interesting difference between the covariant  and
the non-covariant ways of introducing gauge potentials. Namely, the equivalent of the electric vector potential $\vec{C}$ is defined through
\be\label{Ccov}
I\Frs=\mean{\partial \Clc{C}}_v,
\ee
where $\mvec{C}$ is real,
instead of through
\be
\Frs=\mean{\partial \Clc{A}}_v+I\mean{\partial \Clc{C}}_v
\,\,\,{\rm (wrong!)}
\ee
which one might have thought would be the covariant form of (\ref{FrsFo}) and (\ref{Fo}).

Since one can pick only 4 linearly independent paravectors, $\Frs$, like any other biparavector, can be either written as a single product $\mvec{p}\Clc{q}$---and then it is called a simple biparavector--or as a sum of two such products $\mvec{p}\Clc{q}+
\mvec{r}\Clc{s}$. In either case, that decomposition is covariant (unlike the split in electric and magnetic fields).
Given a biparavector $\Frs$ it is easy to check whether it is simple or not: it is simple
if and only if $\Im(\Frs^2)=0$. 
In particular, if a field is describable by the covariant generalization of Euler potentials
(for a discussion of covariant Euler potentials for a different class of fields, and from a very different perspective, see Ref.~\cite{Uchida})
\be\label{FEuler}
\Frs=\mean{\partial\alpha\overline{\partial}\beta}_v,
\ee
it must be simple. Given (\ref{Ccov}) we can introduce covariant electric Euler potentials by
\be\label{IFEuler}
I\Frs=\mean{\partial\lambda\overline{\partial}\mu}_v.
\ee
These two definitions (\ref{FEuler}) and (\ref{IFEuler}) allow us to define field lines in a covariant way, as follows. A 2D surface in spacetime defined by setting $\alpha$ and $\beta$ equal to constants (note we use the paravector $\mvec{r}=t+\vec{r}$ as argument here)
\be\label{abc}
\alpha(\mvec{r})=\alpha_0;\,\,
\beta(\mvec{r})=\beta_0
\ee
 is an invariant surface, if we stipulate that $\alpha$ and $\beta$ are scalar fields, i.e., that they transform under Lorentz transformations as
\be
\alpha'(\mvec{r}')=\alpha(\mvec{r});\,\,
\beta'(\mvec{r}')=\beta(\mvec{r}).
\ee
(And this we substitute on the left-hand sides of (\ref{abc}).)
This invariant surface is swept out over time by a given magnetic field line.
Conversely, a magnetic field line in a given inertial frame of reference is then determined by the intersection of the invariant 2D surface with a spacelike hyperplane, i.e., a hyperplane whose normal is timelike: in other words, it is simply determined by fixing
the time coordinate in the given frame in Eq.~(\ref{abc}).
This then gives the covariant definition of magnetic field lines, and it agrees with that given by Newcomb in \cite{Newcomb}.
For finding electric field lines in a given frame of reference, we go through the same procedure starting with (\ref{IFEuler}), and end up setting
\be\label{lmc}
\lambda(\mvec{r})=\lambda_0;\,\,
\mu(\mvec{r})=\mu_0
\ee
and fixing the time coordinate in that frame.
This procedure then determines what field lines in one reference frame look like in another. Namely, we simply identify magnetic and electric field lines by the pair of constants
$(\alpha_0,\beta_0)$ and $(\lambda_0,\mu_0)$, respectively.

Returning to the definition of helicity density (\ref{heln}) from the previous Section we see it is not covariant, as it is not an alternating product of paravectors and their Clifford conjugates. But it can be turned into a sum of two different covariant terms by inserting an arbitrary field-independent paravector $\mvec{q}$ and its Clifford conjugate (\ref{Clc}), like so:
\be\label{helc}
h(\mvec{q})=\Frs\mvec{q}\Fo^\dag+\Fo^\dag\Clc{q}\Frs=2\mean{\Frs\mvec{q}\Fo^\dag}_s.
\ee
The first term  transforms  as a paravector, the second as the Clifford conjugate of a paravector, and their sum is a scalar.
We recognize now the helicity density (\ref{heln}) as simply proportional to this scalar. For example, by choosing  $\mvec{q}=\mvec{u}:=(1,0,0,0)$ it follows that the helicity density equals $h(\mvec{u})/2$.  
Given this, it is easy to see that the helicity being an integral over all space of the zeroth component of four-vectors is invariant under Lorentz transformations (just like electric charge is, for instance). In fact, this invariance also follows from the expression Eq.~(\ref{Hmab}), which shows the (magnetic) helicity is expressable in terms of the scalar Euler potentials. 

\section{Numerical results}\label{Num}
\subsection{The \HR solutions}
We now apply the results of the preceding Sections to the \HR solutions.
Those solutions are usually given in terms of complex functions $\eta$ and $\zeta$ through
\bea\label{BERanada}
\vec{B}&=&\frac{1}{4\pi i}\frac{1}{(1+|\eta|^2)^2}\grad\eta\times\grad\eta^*,\nonumber\\
\vec{E}&=&\frac{1}{4\pi i}\frac{1}{(1+|\zeta|^2)^2}\grad\zeta\times\grad\zeta^*.
\eea
For completeness, we give them here explicitly
\bea
\eta&=&\frac{Az+t(A-1)+i(tx-Ay)}{Ax+ty+i(A(A-1)-tz)},\nonumber\\
\zeta&=&\frac{Ax+ty+i(Az+t(A-1))}{tx-Ay+i(A(A-1)-tz)},\label{ez}
\eea
with $A=(|\vec{r}|^2-t^2+1)/2$.
But by switching variables from $\eta,\eta^*$ to
$\phi=\arctan(\Re(\eta)/\Im(\eta))$ and $|\eta|^2$ (and similarly for $\zeta$ and $\zeta^*$) we get the (non-covariant) Euler form for the magnetic field with
\be
\alpha=\frac{\phi}{2\pi};\,\,\beta=\frac{1}{1+|\eta|^2},
\ee
and a similar result for the electric field in terms of $\zeta$.
Remarkably, the same functions also can be used as the covariant Euler potentials.

The \HR solutions are known to be null fields, i.e., they satisfy both $\vec{E}\cdot\vec{B}=0$ and $|\vec{E}|^2=|\vec{B}|^2$. We can, therefore, define magnetic and electric field lines covariantly, and both types of field lines move at the speed of light.

The magnetic helicity
of these solutions is obtained through Eq.~(\ref{Hmab}) by noting that
\be
\beta_{\max}-\beta_{\min}=1;\,\,\beta(L)=1;\,\,\alpha_{\max}-\alpha_{\min}=1,
\ee
so that $H_m=1$. The same result is obtained in the same way for the electric helicity, $H_e=1$.
These results had been obtained before in a purely topological way  as the index of the Hopf map in \cite{Ranada1992}.

We consider here two 2D surfaces in 3D space by considering the locus of points where
$\alpha(\vec{r},t_0)=\alpha_0$ at some fixed time $t_0$, and, similarly,
$\beta(\vec{r},t_0)=\beta_0$. These two surfaces are not invariant. One topological property of these surfaces can be immediately obtained from the character of field lines: the field lines constitute a tangential vector field that is nowhere zero. So it has an index 0, and therefore, by the Poincar\'e-Hopf theorem, can only exist on a closed surface if it has Euler characteristic equal to zero. That is, 
the surface, if it is closed, must be topologically equivalent to a torus.
\begin{figure}[h]
\begin{center}
\includegraphics[width=3.3in]{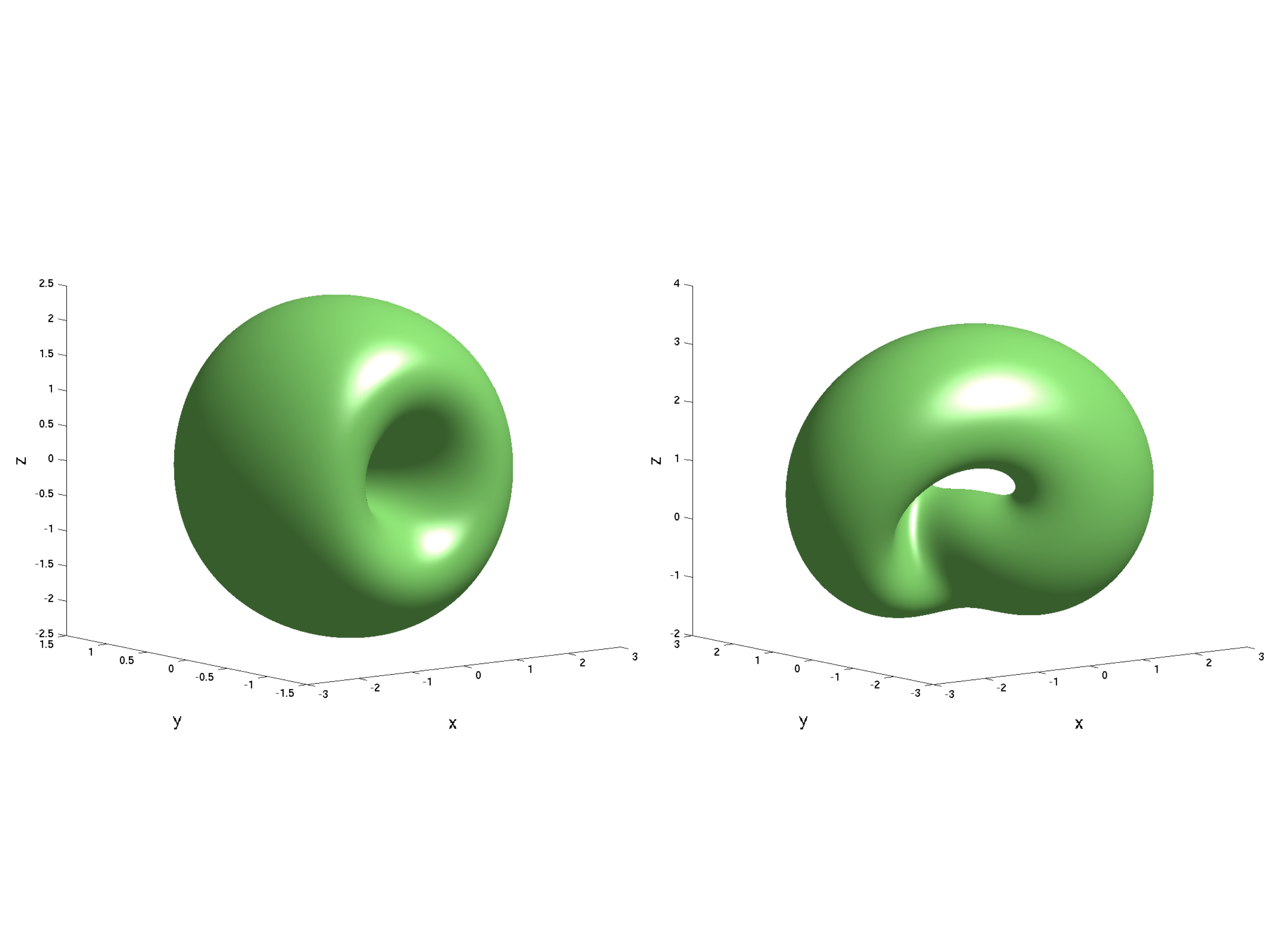}
\caption{Surface $\beta_0=0.5$ at two different times, $t=0$ on the left and $t=1$ on the right, in the lab frame.
Note the axes in left and right plots are not quite the same, and the surface has in fact moved up, i.e., in the positive $z$ direction.}
\label{LabF}
\end{center}
\end{figure}
In Fig.~\ref{LabF} we  plot a particular surface $\beta=$ constant at two different times in the ``lab frame,'' by which we simple mean the frame in which the expressions (\ref{ez}) hold.  We pick here and in all similar figures below the value $\beta_0=1/2$, corresponding to $|\eta|^2=1$. The  topology of the torus is clearly visible.

\begin{figure}[h]
\begin{center}
\includegraphics[width=3.3in]{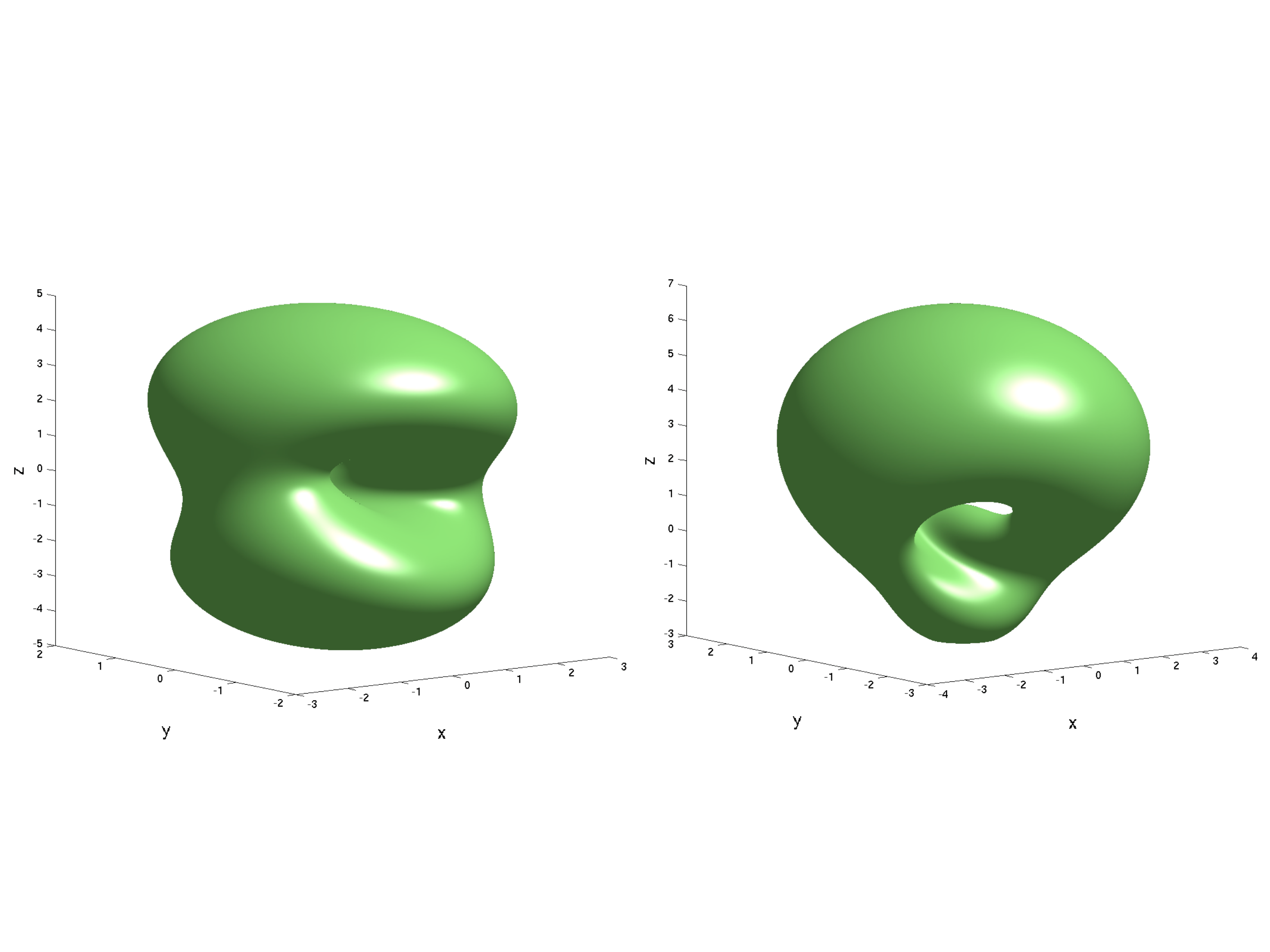}
\caption{Same as Fig.~\ref{LabF} but in a reference frame moving in the positive $z$ direction with speed $c/2$.}
\label{vz+F05}
\end{center}
\end{figure}

\begin{figure}[h]
\begin{center}
\includegraphics[width=3.3in]{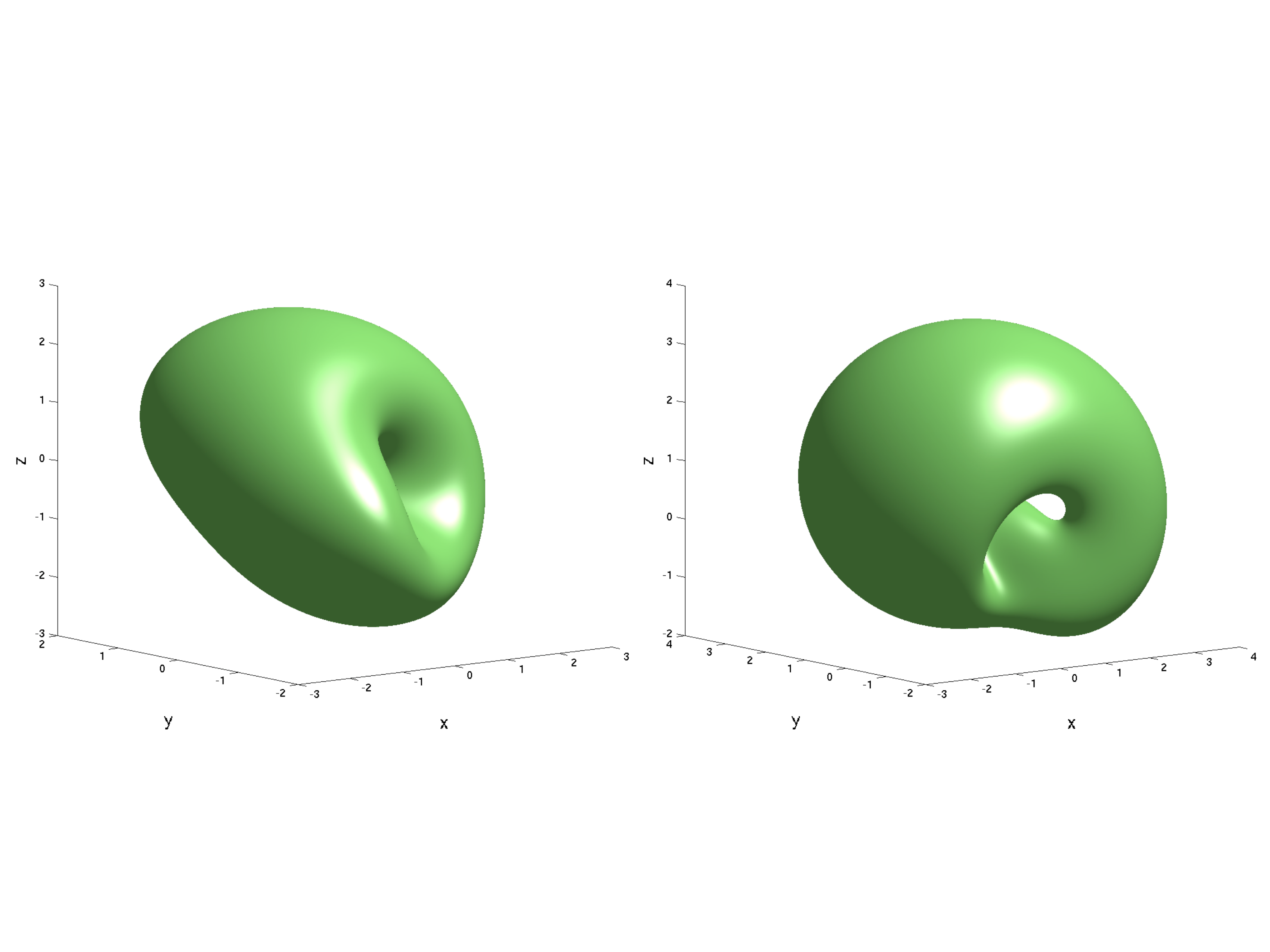}
\caption{Same as Fig.~\ref{LabF} but in a frame moving in the positive $x$ direction with speed $c/2$.}
\label{vx+F05}
\end{center}
\end{figure}
Then we plot the same surfaces
 in two other reference frames, both moving at a speed of $c/2$, in Figs.~\ref{vz+F05} and \ref{vx+F05}. That is, in those references frames we pick $t'=0$ and $t'=1$ (while making the standard assumption that the origins in the different frames coincide at $t=t'=0$), and we keep the constant $\beta_0=1/2$ the same.  The plots confirm that the Lorentz-transformed surfaces stay  closed and topologically equivalent to a torus. We note that
the surfaces are not stationary, so the deformations cannot be explained simply from just length contraction.

\begin{figure}[h]
\begin{center}
\includegraphics[width=3.3in]{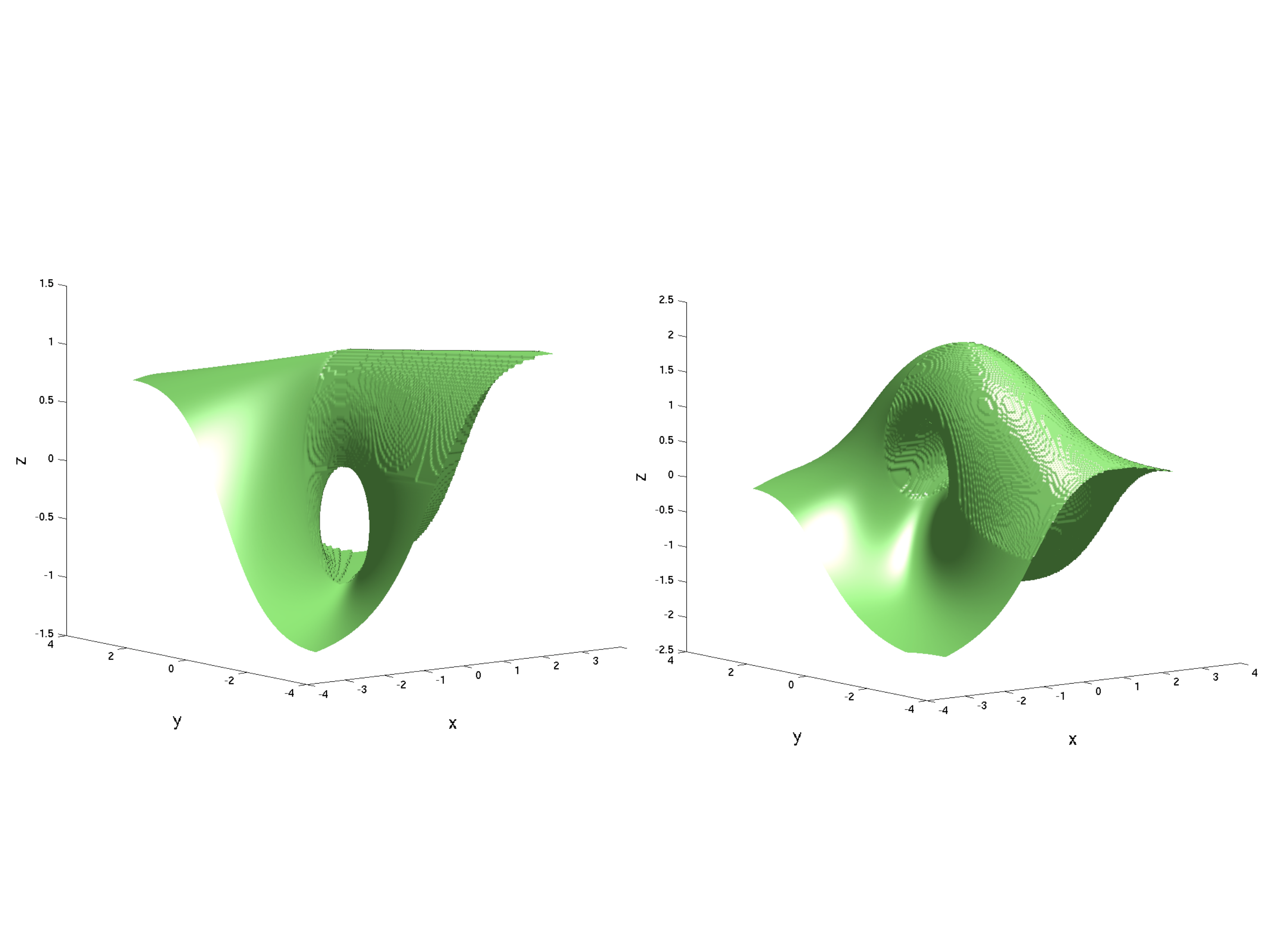}
\caption{Surface $2\pi\alpha_0=0.1$ at two different times, $t=0$ on the left and $t=1$ on the right, in the lab frame.
Note the axes in left and right plots are not quite the same, and the surface has in fact moved up, i.e., in the positive $z$ direction.}
\label{LabG}
\end{center}
\end{figure}

\begin{figure}[h]
\begin{center}
\includegraphics[width=3.3in]{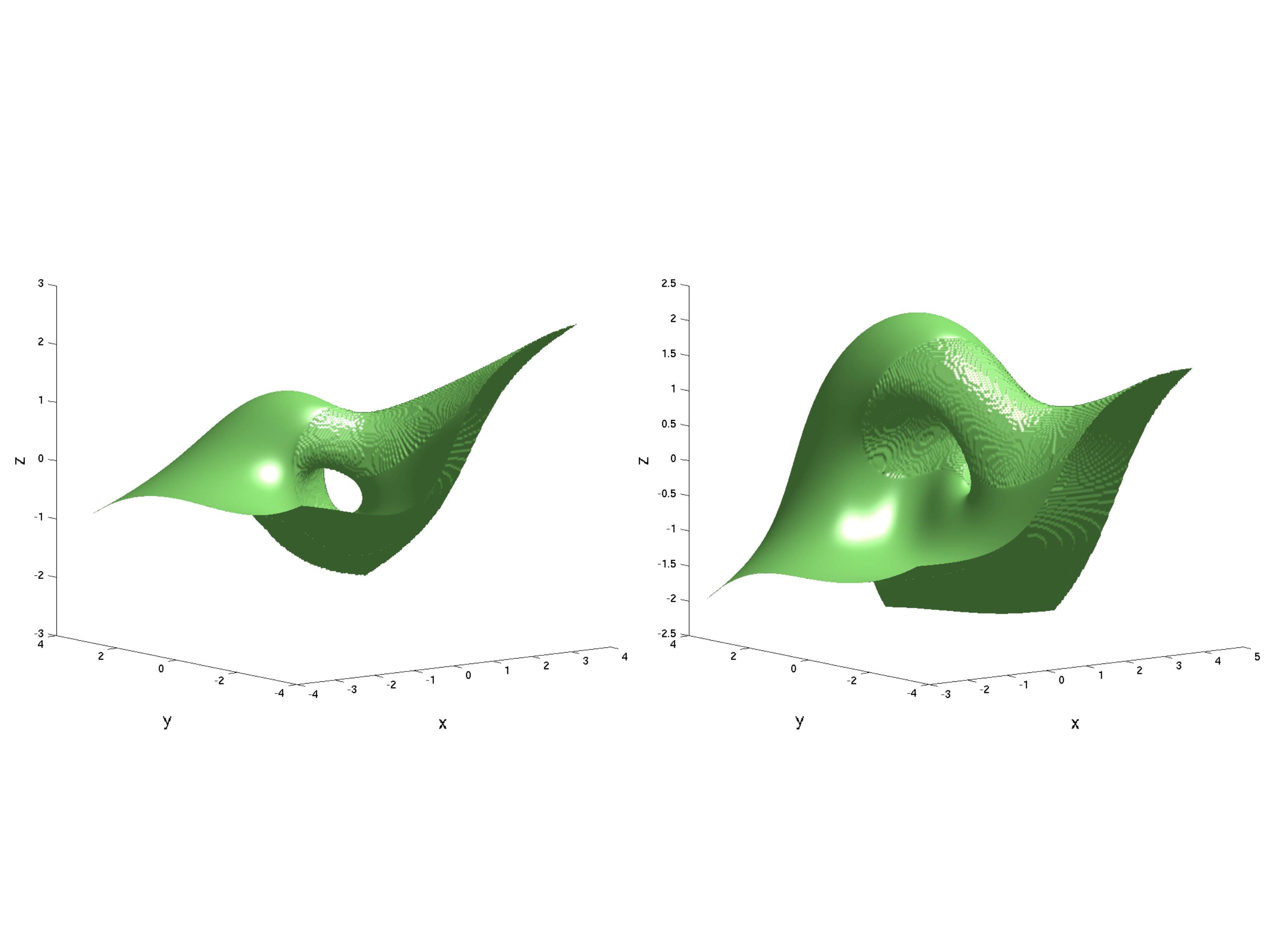}
\caption{Same as Fig.~\ref{LabG} but in a frame moving in the positive $x$ direction with speed $c/2$.}
\label{vxG}
\end{center}
\end{figure}

In Figs.~\ref{LabG}--\ref{vxG} we then plot surfaces $\alpha=\alpha_0$ constant, again in the same three reference frames as before, and at the same times.
The constant is picked the same in the three figures, namely $\alpha_0=0.1/(2\pi)$, so that we can say we plot the same surfaces as seen by different observers.
These surfaces are not closed, but the figures do show
they have one handle.

\begin{figure}[h]
\begin{center}
\includegraphics[width=3.3in]{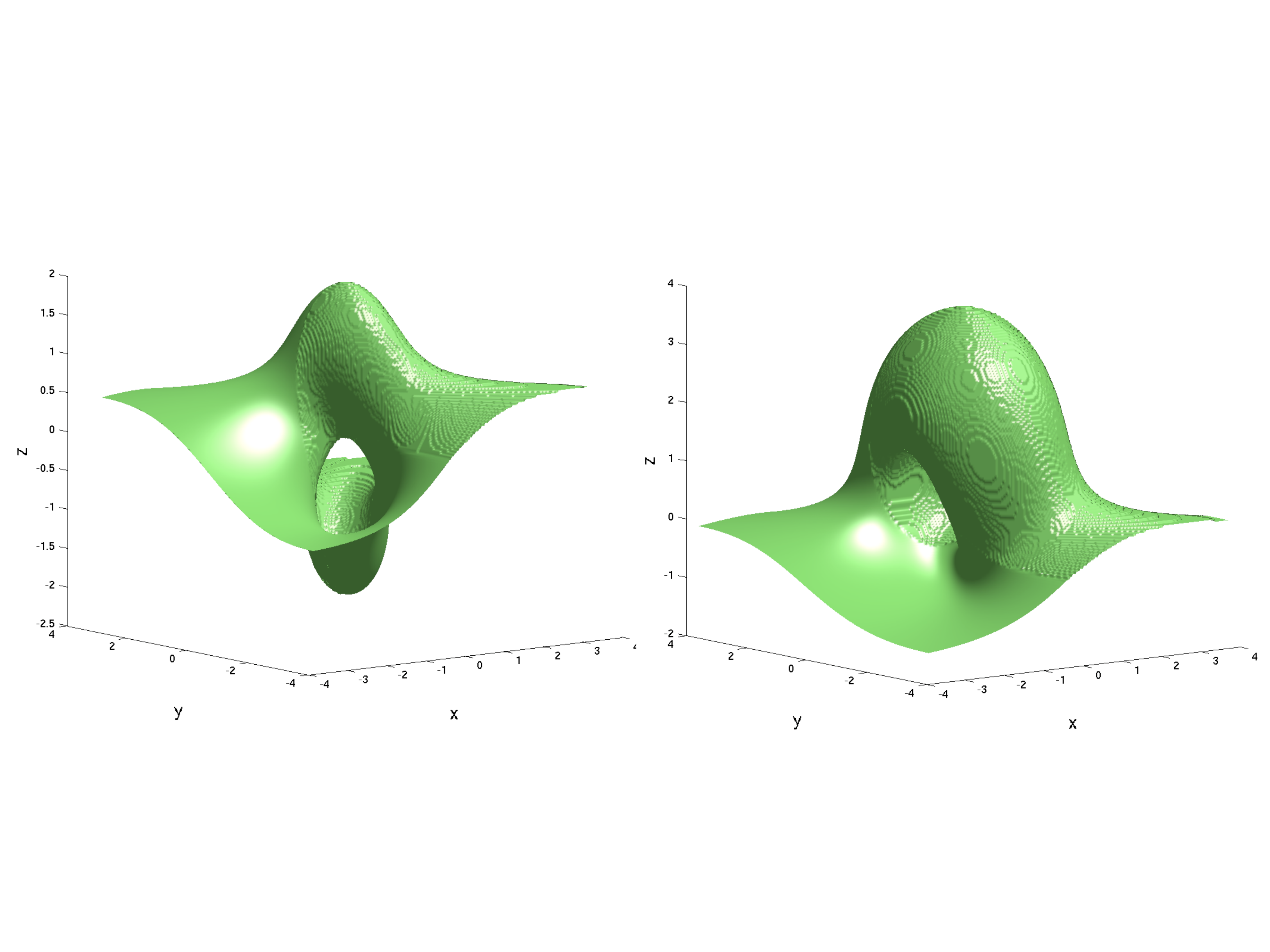}
\caption{Same as Fig.~\ref{LabG} but in a reference frame moving in the positive $z$ direction with speed $c/2$. }
\end{center}
\end{figure}

\begin{figure}[h]
\begin{center}
\includegraphics[width=3.3in]{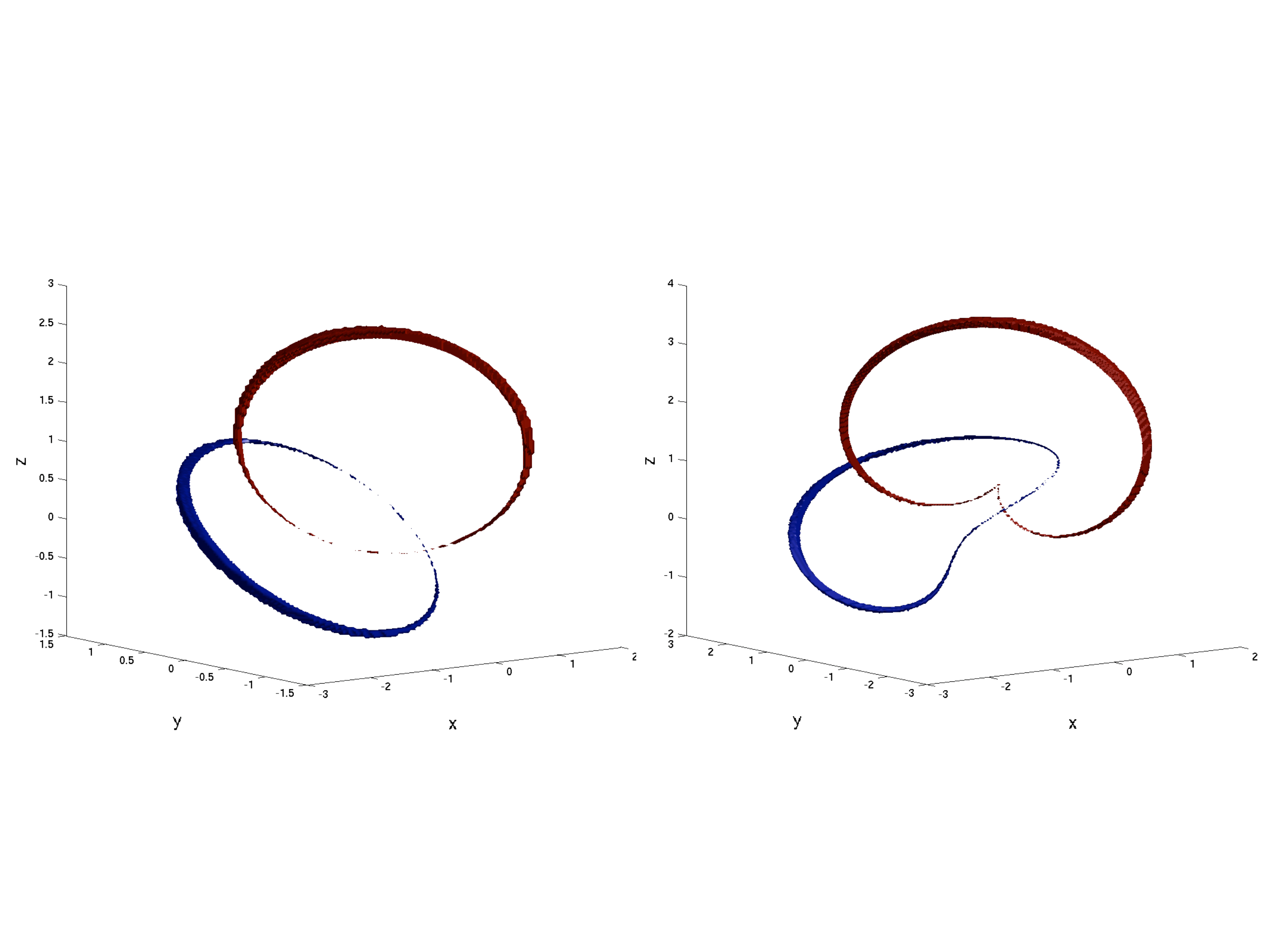}
\caption{In blue: magnetic field lines in the lab frame obtained as intersections of the surfaces plotted in Figs.~\ref{LabF} and \ref{LabG}. The other field line (red) is electric, and is obtained with the same procedure applied to the surfaces $\lambda=\lambda_0=0.1/(2\pi)$   and $\mu=\mu_0=0.5$.
On the left $t=0$, on the right $t=1$.}
\label{LabL}
\end{center}
\end{figure}

\begin{figure}[h]
\begin{center}
\includegraphics[width=3.3in]{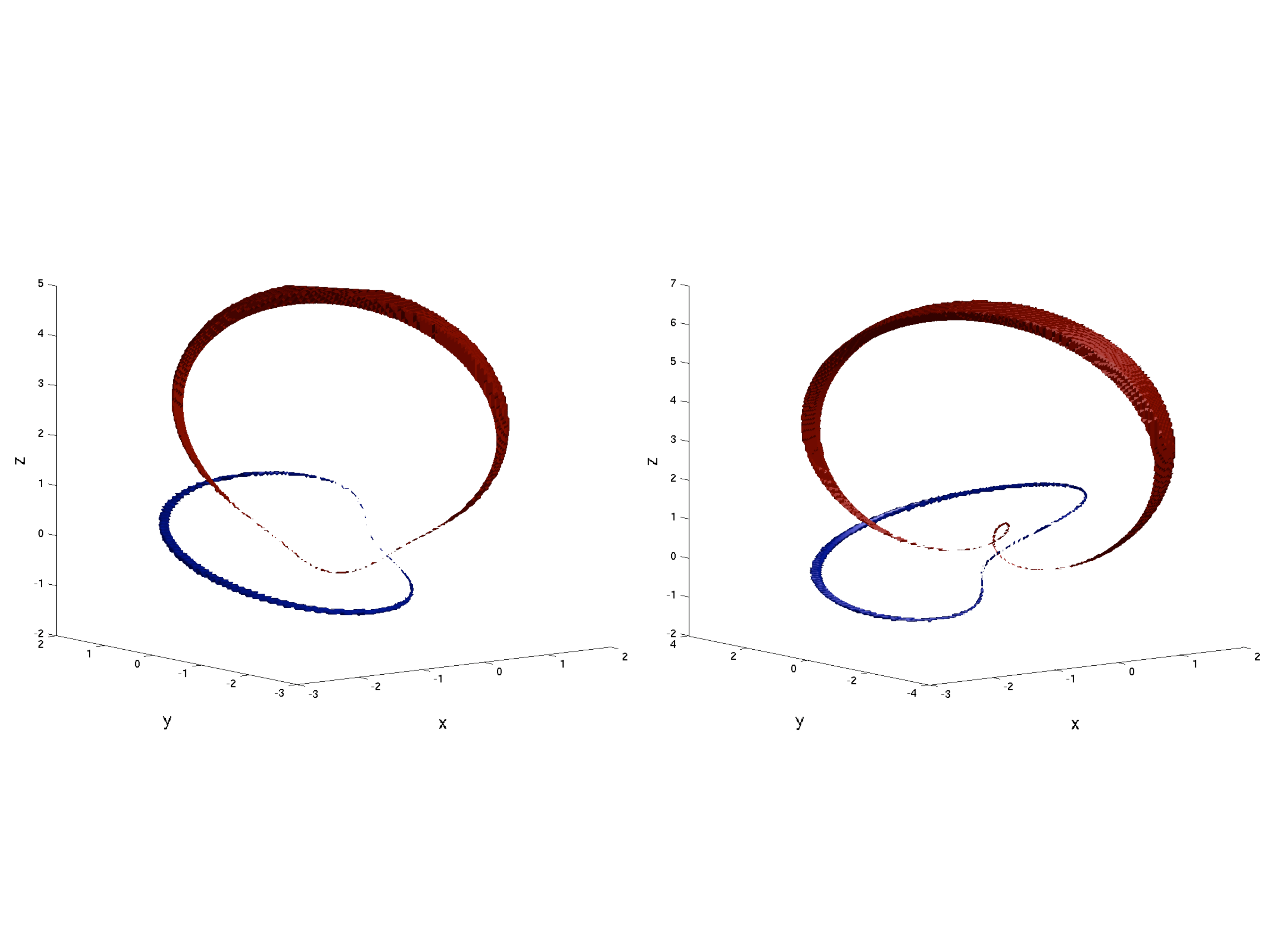}
\caption{Same as Fig.~\ref{LabL} but in a reference frame moving in the positive $z$ direction with speed $c/2$.}
\label{vz+L}
\end{center}
\end{figure}

\begin{figure}[h]
\begin{center}
\includegraphics[width=3.3in]{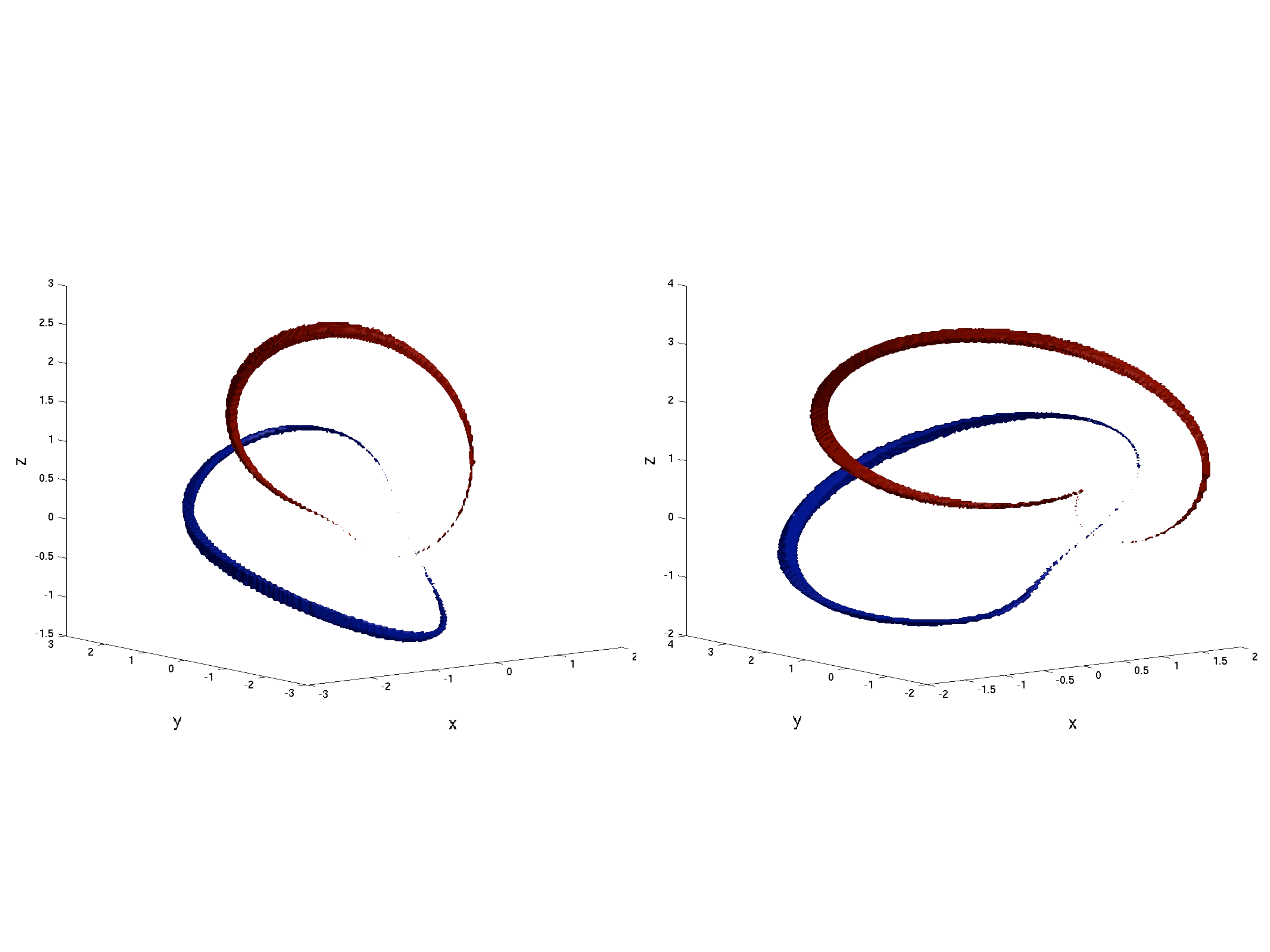}
\caption{Same as Fig.~\ref{LabL} but in a frame moving in the positive $x$ direction with speed $c/2$.}
\label{vx+L}
\end{center}
\end{figure}

In the last set of three figures, \ref{LabL}, \ref{vz+L} and \ref{vx+L} we plot the intersections of the surfaces shown in the previous six figures. That is,
we plot the magnetic field lines corresponding to the two constants $\beta_0=0.5$ and $2\pi\alpha_0=0.1$.
In addition we plot in the same figures the electric field line corresponding to the intersection of surfaces $\lambda=\lambda_0$ and $\mu=\mu_0$ with the constants chosen the same as for the magnetic case, i.e.,
$\lambda_0=\alpha_0$ and $\mu_0=\beta_0$.

\subsection{Extension of \HR solutions}
Ran\~ada has given an extension of his solutions. Namely, given the complex function $\eta$, we can pick, according to \cite{Ranada1997} the following pair of Euler potentials
\bea\label{new}
\alpha_n&=&\frac{1}{2\pi}\arctan \left(\frac{\Re(\eta^n)}{\Im(\eta^n)}\right)\nonumber\\
\beta&=&\frac{1}{1+|\eta|^2}
\eea
for any integer $n$. 
That is, the function $\eta\equiv
|\eta|\exp(i\phi)$ in (\ref{BERanada}) is replaced by the function
$\eta_n=|\eta|\exp(ni\phi)$.

But these solutions are really the same: after all, setting $\eta_n$ equal to a (complex) constant is equivalent to setting $\eta$ equal to a (different, complex) constant, and so the field lines of $\eta$ and $\eta_n$ are the same. 
Ra\~nada does state that the magnetic helicity changes by a factor of $n^2$, but that arises trivially from the simple relations $\vec{B}_n=n\vec{B}_1$ and $
\vec{A}_n=n\vec{A}_1$.

\begin{figure}[h]
\begin{center}
\includegraphics[width=2.2in]{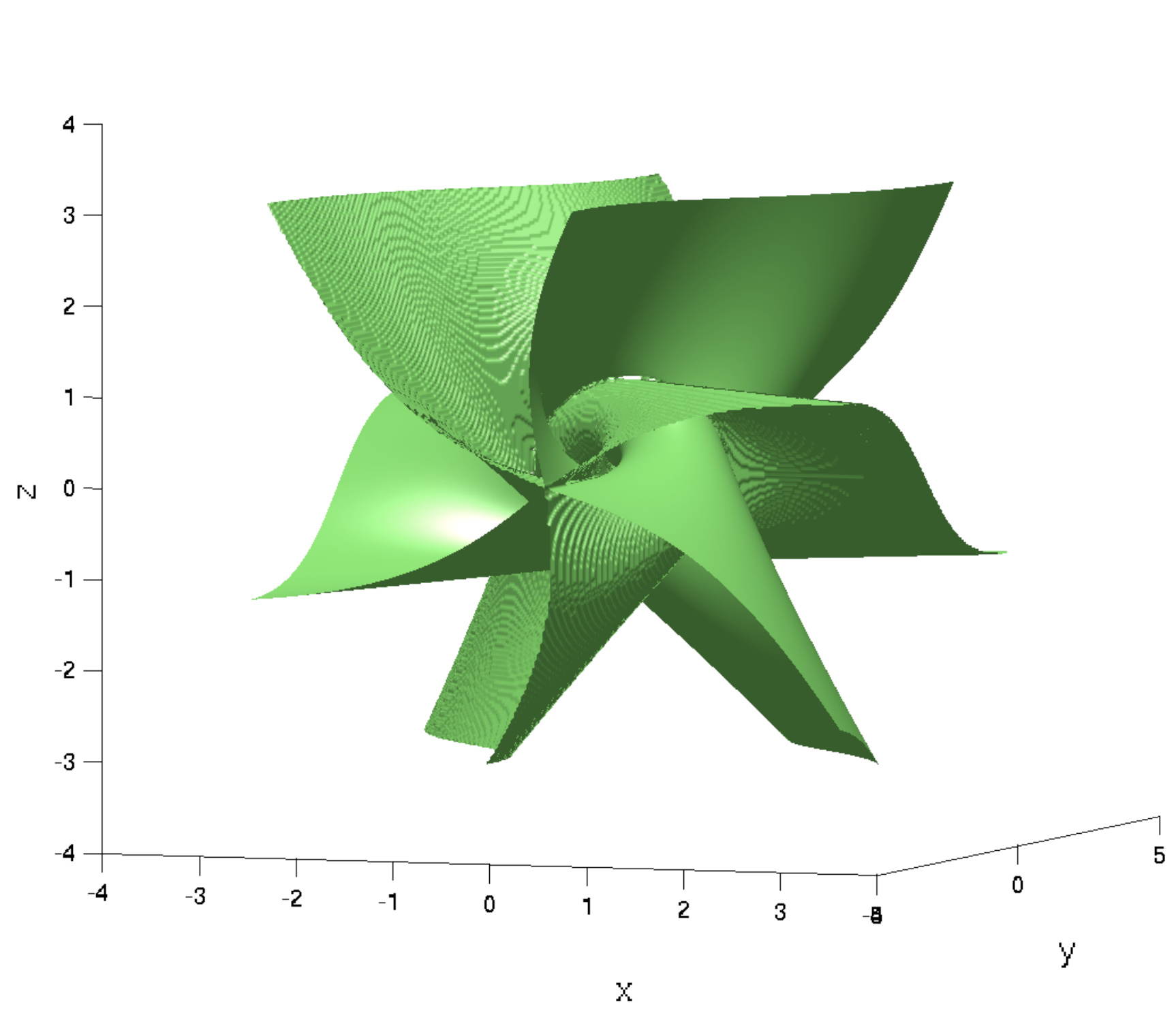}
\caption{Plot of surface of constant $\alpha_3=0.3/(2\pi)$ at time $t=0$ in the lab frame. The multi-valued function $\alpha_3$ has 3 branches, which may intersect at most in points where the gradient is singular. 
The viewing angle is different here than in previous figures, so as to make the structure of the corresponding field line(s) more clearly visible (see next figure).}
\label{3}
\end{center}
\end{figure}

\begin{figure}[h]
\begin{center}
\includegraphics[width=2.2in]{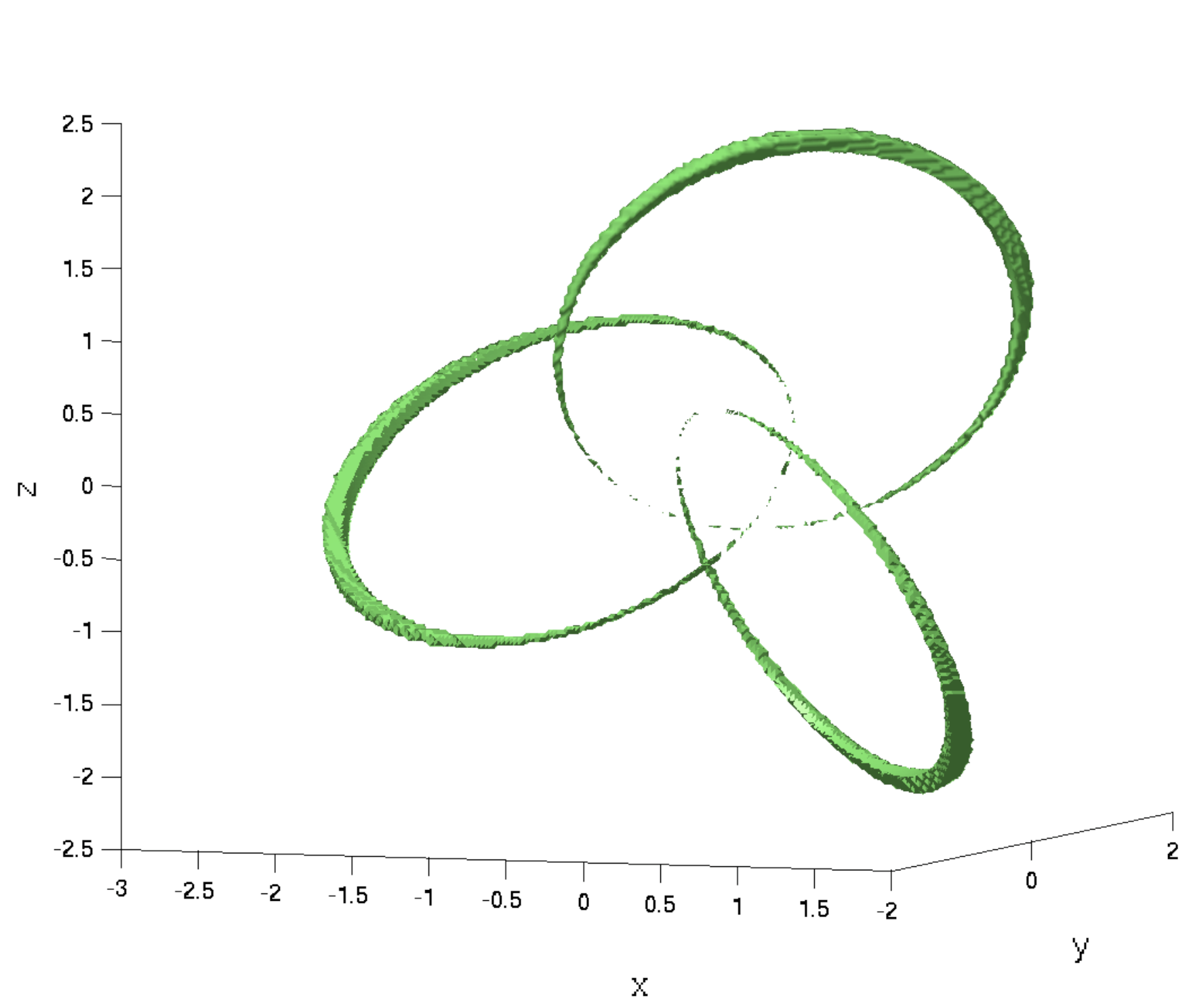}
\caption{Magnetic field lines  as determined by the intersection of the surface plotted in the preceding figure with the surface $\beta=0.5$.}
\label{v0t0G33FL}
\end{center}
\end{figure}

Whereas surfaces $\alpha=$constant are manifolds (even if not simply connected), 
surfaces $\alpha_n=$constant for $n>1$ are not, and consist of $n$ branches. We plot an example in Fig.~\ref{3}, where we chose $n=3$ at time $t=0$, and the constant equals $0.3/(2\pi)$. Three branches are clearly visible.
These three branches correspond to three different surfaces obtained by setting $\alpha$ equal to three constants $\alpha_0=0.1/(2\pi)$ and $\alpha_0\pm 1/3$.
Note that these branches may intersect at most at the points where $|\eta|^2=0$, the branch cut of the function $\alpha_n$ (here, the $x$ axis). One can see in the plot that at those points the gradient of $\alpha_n$ is singular. The field line obtained by intersecting the branches with the surface $\beta=0.5$ is plotted in Fig.~\ref{v0t0G33FL}. That plot shows that this is not a single field line, but three field lines.


\section{The Bateman solutions}\label{Bateman}
In his book from 1915 \footnote{The references to the Aether theory in this book are, with hindsight, amusing and puzzling, since the book was written 10 years after 1905. Those references become more understandable after reading \cite{Earman}}, Bateman constructs a broad class of solutions to the free Maxwell equations that all satisfy the null property, $\Frs^2=0$. Recall that for such solutions electric and magnetic helicities are conserved, and field lines can be defined covariantly.
The description of those solutions in the book comes tantalizingly close to the description used in the present paper, with one important difference: Bateman uses {\em complex} Euler potentials, instead of real ones. This does negate their advantage in finding field lines or expressing the helicity.
The complex solutions satisfy an additional constraint, which may help in constructing real Euler potentials, but we have not succeeded in doing so. 
Even though Bateman's solutions are not given in covariant form in his book, it is straightforward to convert to covariant notation.

In more detail then, Bateman finds pairs of complex functions $\phi_1$ and $\phi_2$, such that (using the {\em lower} of the ambiguous sign he uses in his Eqs.~(2) and (10))
\be
\Frs=\mean{\partial\phi_1\overline{\partial}\phi_2}_v.
\ee
Moreover, these functions satisfy the condition
\be
\mean{\overline{\partial}\phi_1\partial\phi_2}_v=0.
\ee
This constraint  implies $\Frs^2=0$, as can be easily checked within the formalism of GA.

Examples of pairs of functions $\phi_{1,2}$ can be obtained by choosing $\phi_1=\beta_B$ and $\phi_2=\alpha_B/2$, where $\alpha_B$ and $\beta_B$ (conforming to the notation Bateman uses) must be picked with the {\em lower} sign in his Eqs.~(11) or (13)
\bea\label{B1}
\phi_1&=&x\sin\vartheta-y\cos\vartheta-t\nonumber\\
\phi_2&=&\half(x\cos\vartheta+y\sin\vartheta+iz),
\eea
where $\vartheta$ is a parameter that can be chosen arbitrarily,
or
\bea
\phi_1&=&r-t,\nonumber\\
\phi_2&=&\frac{x+iy}{2(z+r)}.
\eea
The \HR solutions are, in fact, a special example of the Bateman solutions, as has been shown in \cite{Besieris}.
Even this correspondence, where both \HR and Bateman solutions are explicitly known, is not easy to demonstrate explicitly.

\section{Summary}
We described a known set of solutions to the free Maxwell equations, the \HR linked and knotted fields,  in a compact geometric way by using Geometric Algebra.
In particular, we represented the solutions in covariant form as
\be\label{Esum}
\Frs=\mean{\partial\alpha\overline{\partial}\beta}_v,
\ee
with $\alpha$ and $\beta$ real scalar potentials.
This way of writing yields {\em magnetic} field lines as the intersections of two such 2D surfaces $\alpha=\alpha_0$ and $\beta=\beta_0$. We plotted examples of these 2D surfaces and their Lorentz-transformed versions, confirming that such surfaces must have genus equal to 1.
Moreover, this description shows that whereas ordinarily field lines cannot be linked, they can when $\alpha$ is multi-valued and its gradient is singular.
The {\em electric} field lines are obtained by writing, similarly,
\be
I\Frs=\mean{\partial\lambda\overline{\partial}\mu}_v,
\ee
with $I$ the unit pseudoscalar volume element (so that the left-hand side corresponds to the dual of $\Frs$).
For the \HR solutions, we showed explicitly that electric and magnetic helicities are individually conserved in time; that helicity is invariant under Lorentz transformations, and that their chirality vanishes, all as a consequence of their ``null property'' $\Frs^2=0$.
Finally, we showed that Bateman's solutions, which have the null property as well, can be written in the form
\be
\Frs=\mean{\partial\phi_1\overline{\partial}\phi_2}_v,
\ee
in terms of {\em complex} potentials that satisfy
\be
\mean{\overline{\partial}\phi_1\partial\phi_2}_v=0.
\ee
The question how these solutions can be rewritten in the form (\ref{Esum}) remains open.
\section*{Acknowledgements}
I thank Dirk Bouwmeester 
for his hospitality and discussions on the \HR solutions and Geometric Algebra with his group during a 6-weeks' visit to Leiden University  
\bibliography{fieldlines}
 \end{document}